\documentclass[a4paper,11pt]{article}
\usepackage{pos}

\title{Thermodynamics with Möbius domain wall fermions near physical point II}

\author[a]{Sinya Aoki}
\author[b]{Yasumichi Aoki}
\author[c]{Hidenori Fukaya}
\author[b]{Jishnu Goswami}
\author[d,e]{Shoji Hashimoto}
\author*[b]{Issaku Kanamori}
\author[d,e,f]{Takashi Kaneko}
\author[b]{Yu Zhang}

\affiliation[a]{Center for Gravitational Physics, Yukawa Institute for Theoretical Physics, Kyoto University, Kyoto 650-0047, Japan}

\affiliation[b]{%
 RIKEN Center for Computational Science (R-CCS),
Kobe 650-0047, Japan}

\affiliation[c]{%
Department of Physics, Osaka University, Toyonaka, Osaka 560-0043, Japan
}

\affiliation[d]{KEK Theory Center, High Energy Accelerator Research Organization (KEK), 
Tsukuba 305-0801, Japan}

\affiliation[e]{School of High Energy Accelerator Science, Graduate University for Advanced Studied (SOKENDAI),\\
Tsukuba 305-0801, Japan}

\affiliation[f]{Kobayashi-Maskawa Institute for the Origin of Particles and the Universe, Nagoya University,\\
 Aichi, 464-8603, Japan}

\emailAdd{kanamori-i@riken.jp}

\abstract{%
We report on our finite temperature 2+1 flavor lattice QCD simulation to
study the thermodynamic properties of QCD near the (pseudo) critical
point employing $N_T=12$ and $16$. The simulation points are chosen
along the lines of constant physics.  The quark mass for
M\"obius domain-wall fermion are tuned by taking
into account the residual mass either by fine-tuning the input quark masses or by post-process using reweighting.
In this talk, we focus on simulation details and present some preliminary results.
}

\FullConference{%
The 39th International Symposium on Lattice Field Theory,\\
8th-13th August, 2022,\\
Rheinische Friedrich-Wilhelms-Universität Bonn, Bonn, Germany
}

\begin{document}
\maketitle

\section{Introduction}
Since the first physical point, continuum limit study was reported on the finite temperature phase transition of 2+1 flavor QCD being a crossover \cite{Aoki:2006we}, lots of studies have provided consistent results to their result.
However, recent 3 flavor studies suggest that one should reexamine the result
with different lattice fermions and/or finer lattices \cite{Jin:2014hea,Jin:2017jjp,Bazavov:2017xul,Kuramashi:2020meg,Dini:2021hug,Cuteri:2021ikv}.
The common wisdom of the 3 flavor system
was: the transition is first order near the chiral limit and becomes crossover as the fermion mass becomes larger, is now under serious investigation.
As the finer lattice simulation becomes available, 
the critical mass between the first order and the crossover region becomes smaller.
It also changes with the lattice fermion formulation used in the simulation.
It is almost obvious that controlling lattice artifact in this study is quite a challenging task. 
It motivates us to reexamine the quark mass dependence of the finite
temperature transition depicted in the famous Columbia plot.

In this study, we use M\"obius domain-wall fermion that has almost exact
chiral symmetry and investigate the light quark mass dependence at the physical strange quark mass.
As the spontaneously breaking of the chiral symmetry characterizes
the finite temperature transition, using a fermion formulation with chiral symmetry is very important, especially the system is close to the chiral limit.
We use the same fermion formulation used in our zero temperature simulation,
of which details are summarized in the supplemental material in \cite{Colquhoun:2022atw}.
We choose the simulation points along the line of constant physics (LCP) at
which the light quark mass is fixed in the physical unit.
The temperature is controlled through changing the lattice spacing $a$.
To this end, we utilize the zero temperature simulation results to
obtain the gauge coupling $\beta$ dependence of the lattice spacing and 
the renormalization factor of the quark mass.
The details of the $\beta$-dependence and the choice of lattice parameters
are presented in this conference by one of the authors (Y.A.) \cite{talk_yaoki}.

In this article, we first discuss at which quark mass we should simulate
and the effect of the additive residual mass correction.
The input quark mass must be tuned to correct the residual mass effect
for the coarse lattice we use $N_T=12$, but the mass reweighting also works
for the fine lattice with $N_T=16$. 
We present out preliminary results on chiral condensate and disconnected susceptibilities as well as the topological susceptibility.

\section{Simulation Setup}

\begin{figure}
 \centering
\includegraphics[width=0.48\linewidth]{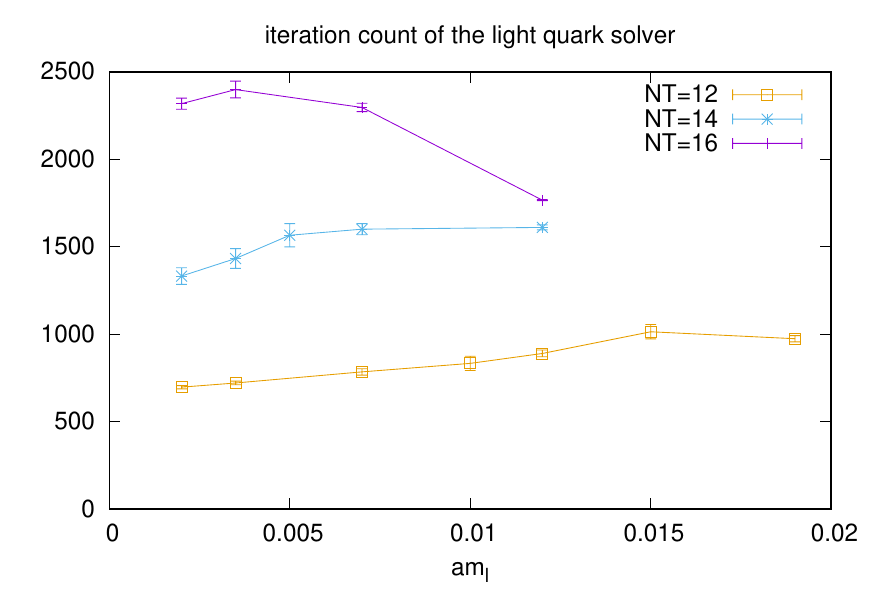}
 \hfil
\includegraphics[width=0.48\linewidth]{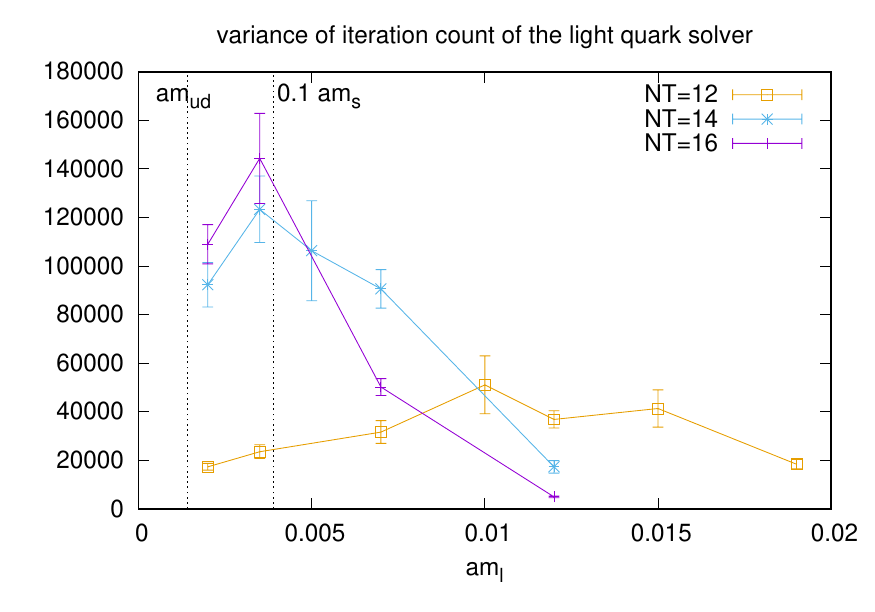}
\caption{Iteration counts of the light quark solver during HMC (left panel)
and the variance (right panel) with fixed lattice spacing $1/a\simeq 1.8$ GeV.
The temperature is $T=153$ MeV ($N_T=16$), $175$ MeV ($N_T=14$) and $205$ MeV ($N_T=12$).}

\label{fig:iter_l_fixed_T}
\end{figure}

\begin{figure}
\centering
\raisebox{2em}{%
\includegraphics[width=0.4\linewidth]{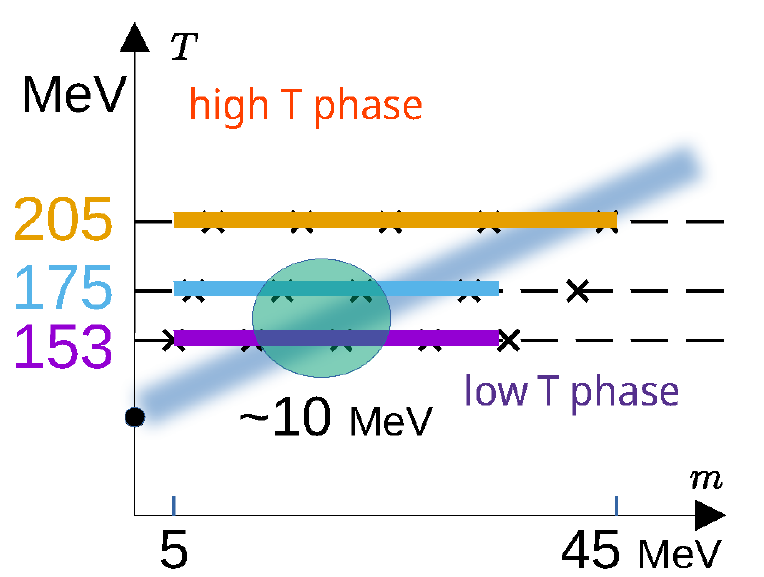}
}
\hfil
\includegraphics[width=0.45\linewidth]{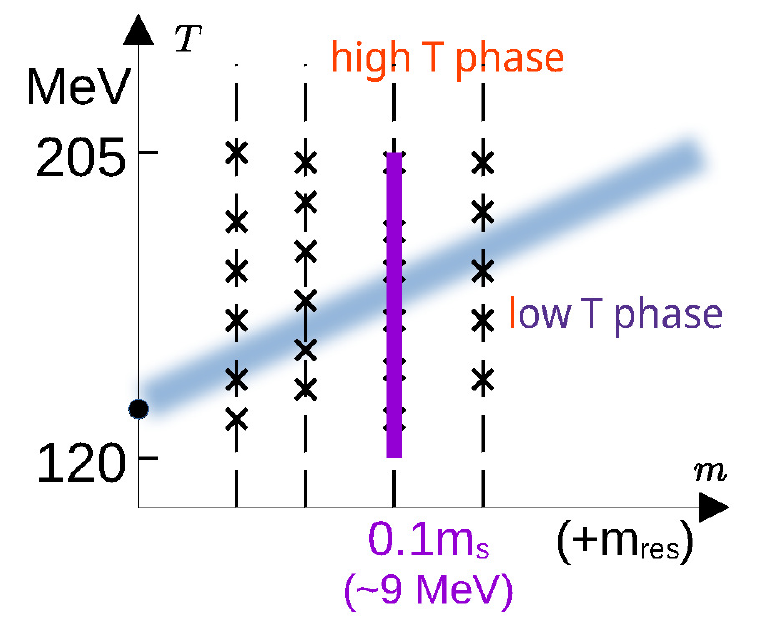}

\caption{Sketch of $m$-$T$ parameter space.  Cross ($\times$) symbols represent simulation points.  The left panel is fixed temperature scanning, and the right panel is for the Line of Constant Physics (LCP) with the fixed light quark mass.}
\label{fig:m-T_scan}
\end{figure}

We first scanned $m$-$T$ parameter space along fixed temperature lines
to find the parameter region to prepare LCP configurations.
We measured several gluonic observables: plaquette, topological charge, Polyakov loop
and their susceptibilities.
We also monitored the iteration counts and the variance
of the light quark solver during the molecular dynamics of HMC.
The iteration count is not a physical observable but it is sensitive to
the spectrum of the Dirac operator.
It turned out that the variance of the iteration counts exhibits
most clearly the sign of (pseudo) phase transition (Fig. \ref{fig:iter_l_fixed_T}).
The figure shows that the peak of the variance is around $m_l \sim 10$ MeV.
From this information we can sketch the phase diagram in $m$-$T$ plain as 
the left panel of Fig.~\ref{fig:m-T_scan} (the figure assumes crossover for $m_l>0$).
Since the quark mass explicitly breaks the chiral symmetry, we expect that
the heavier light quark pushes the upper limit of the temperature of broken phase as in the figure.
We start with the light quark mass $m_l=0.1 m_s$ ($\simeq 9$ MeV)
 before studying the physical point and the chiral limit in the end.  To cover the (pseudo) phase transition 
in this range of light quark mass, we set the temperature 
in $120$ MeV $\lesssim T \lesssim 205$ MeV (right panel of Fig.~\ref{fig:m-T_scan}).

\begin{figure}
 \centering
\includegraphics[width=0.55\linewidth]{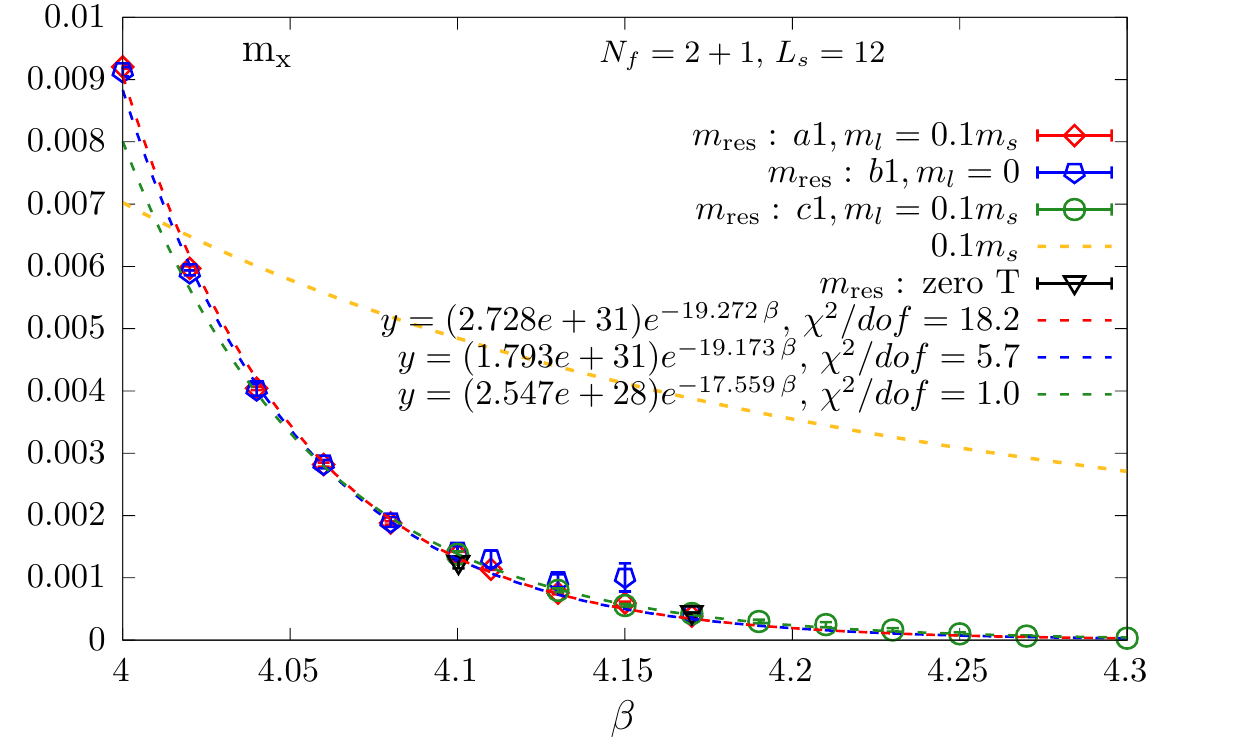}
\caption{Residual mass in lattice unit.
The lattice size is $24^3 \times 12$ (a1, b1) and $32^3\times 16$ (c1).
The coarse ensembles ($N_T=12$) use $4.00 \leq \beta \leq 4.17$ and the 
fine ones ($N_T=16$) use $4.10 \leq \beta \leq 4.30$.
Red, blue and green dashed lines are fits to a1, b1 and c1 data, respectively.}
\label{fig:mres}
\end{figure}

Since we use a finite 5th-dimensional extension $L_s=12$ and finite lattice spacing in the simulations,
the chiral symmetry is broken and the quark mass receives an additive correction.
The size of the correction $m_{\mathrm{res}}$ is
\begin{equation}
  m_{\mathrm{res}}
 = R(t) = \frac{\sum_{\vec{x}} \langle J_{5q}(\vec{x},t)P(\vec{0},0)\rangle}{\sum_{\vec{x}}\langle P(\vec{x},0)P(\vec{0},0)\rangle},
\end{equation}
where $J_{5q}$ is the pseudo scalar density at the mid point in the 5th extent.
As emphasized in the talk by Y.A., it is crucially important to
take into account the effect of $m_{\mathrm{res}}$ 
in the LCP simulation \cite{talk_yaoki}.
Figure~\ref{fig:mres} shows the $\beta$-dependence of $m_{\mathrm{res}}$
measured by using several parameter sets.
The plot shows almost no dependence on the input bare quark mass or the lattice volume and data points are well described by an exponential ansatz in $\beta$.
For the coarse lattice ($N_T=12$), which uses $4.00 \leq \beta \leq 4.17$,
the size of $m_{\mathrm{res}}$ is compatible or even larger than $0.1 m_s$ plotted in yellow dashed line.
We therefore need to shift the input mass to $m_l - m_{\mathrm{res}}$ to 
cancel the effect of $m_{\mathrm{res}}$.
The fine lattice ($N_T=16$) uses larger $\beta$, $4.10\leq\beta \leq 4.30$,
and the correction is small compared to the target light quark mass $0.1m_s$.
We therefore use mass reweighting on the configurations generated without
correcting $m_{\mathrm{res}}$ effect.
To determine the value of $m_{\mathrm{res}}$, we combine the measured values and 
results from a fit with exponential ansatz.
An exponential fit by using $N_T=16$ data has a good $\chi^2$ value so we determine the $m_{\mathrm{res}}$ for $\beta\geq 4.10$ from the fit result, which has smoother $\beta$-dependence than the measured value itself (the bottom fit in the Figure).
On the other hand, exponential fits with $N_T=12$ data have rather poor $\chi^2$ value so we use the measured values to determined the $m_{\mathrm{res}}$ for $\beta<4.10$

\section{Results}

\begin{figure}
 \centering
\includegraphics[width=0.47\linewidth]{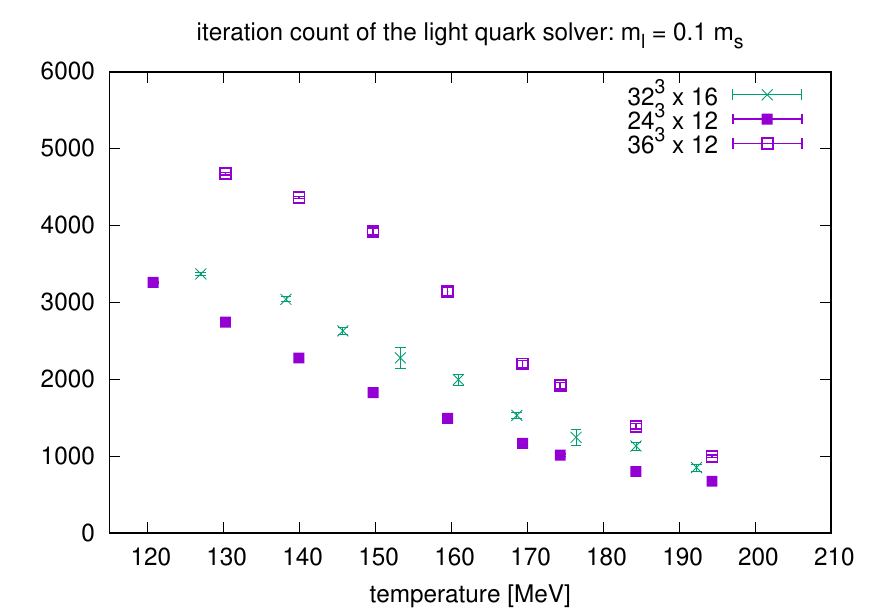}
\hfil
\includegraphics[width=0.47\linewidth]{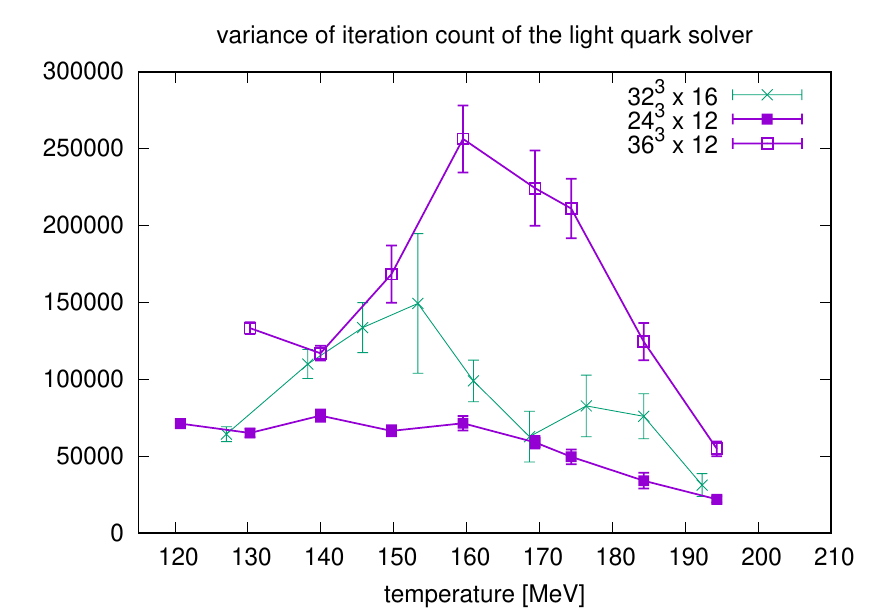}
\caption{
Iteration counts of CG solver for the light quark during HMC (left panel)
and the variance (right panel).  
Finer lattice data ($32^3 \times 16$, green cross) has been reweighted from the
configurations without $m_{\mathrm{res}}$ corrections.}
\label{fig:iter_l}
\end{figure}

\begin{figure}
\noindent
\centering
\includegraphics[width=0.33\linewidth]{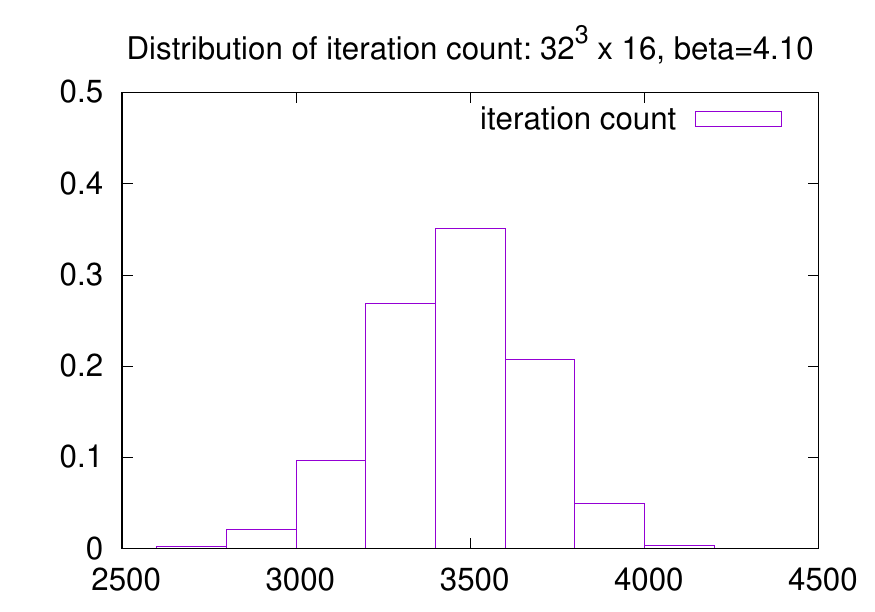}%
\hfil
\includegraphics[width=0.33\linewidth]{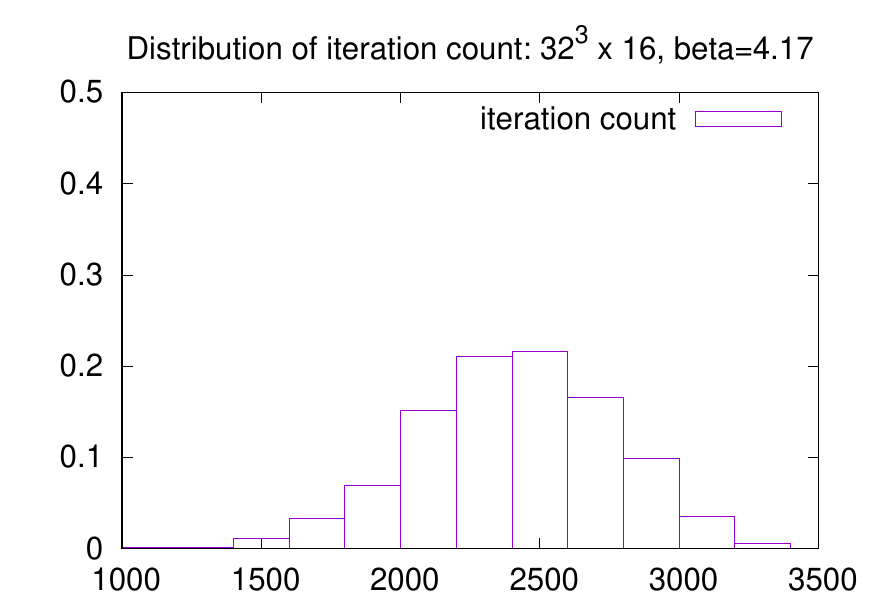}%
\hfil
\includegraphics[width=0.33\linewidth]{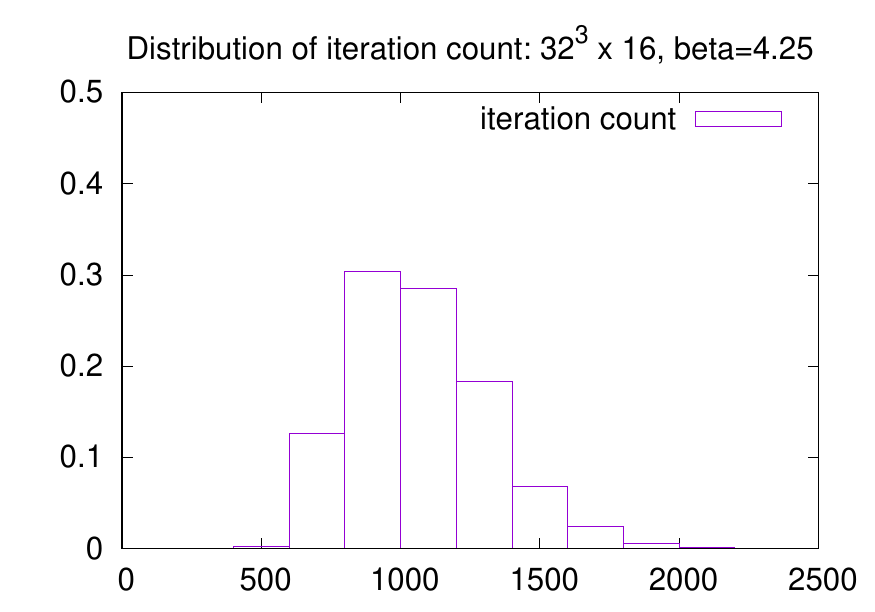}

\noindent
\centering
\includegraphics[width=0.33\linewidth]{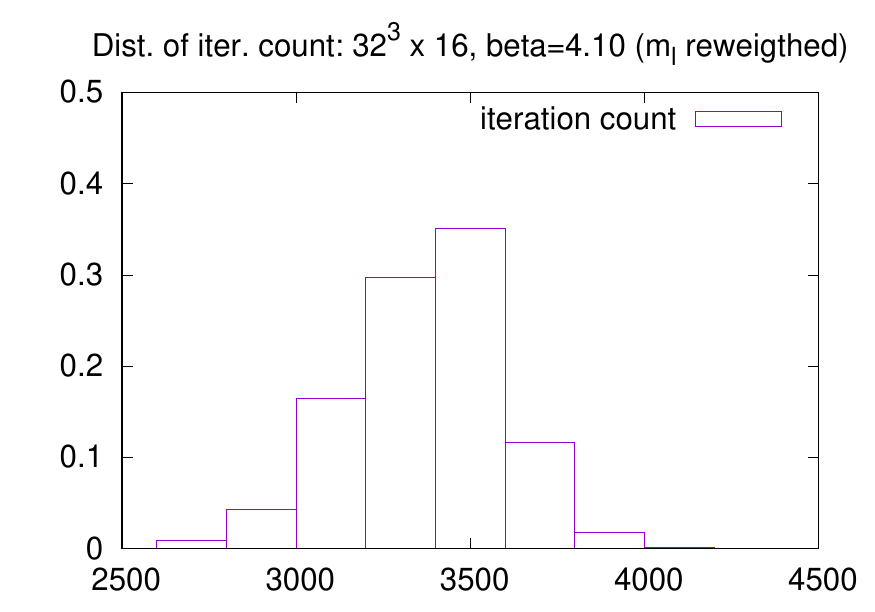}%
\hfil
\includegraphics[width=0.33\linewidth]{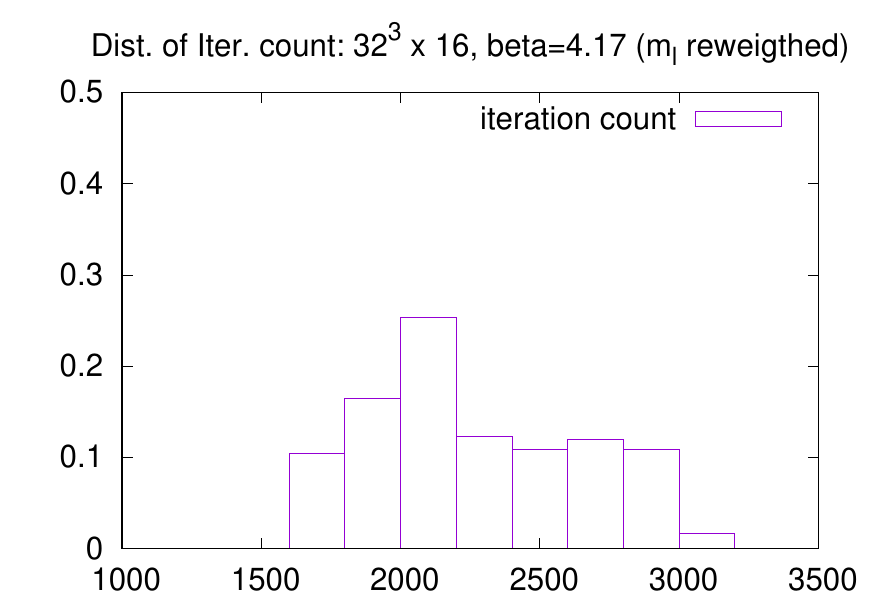}%
\hfil
\includegraphics[width=0.33\linewidth]{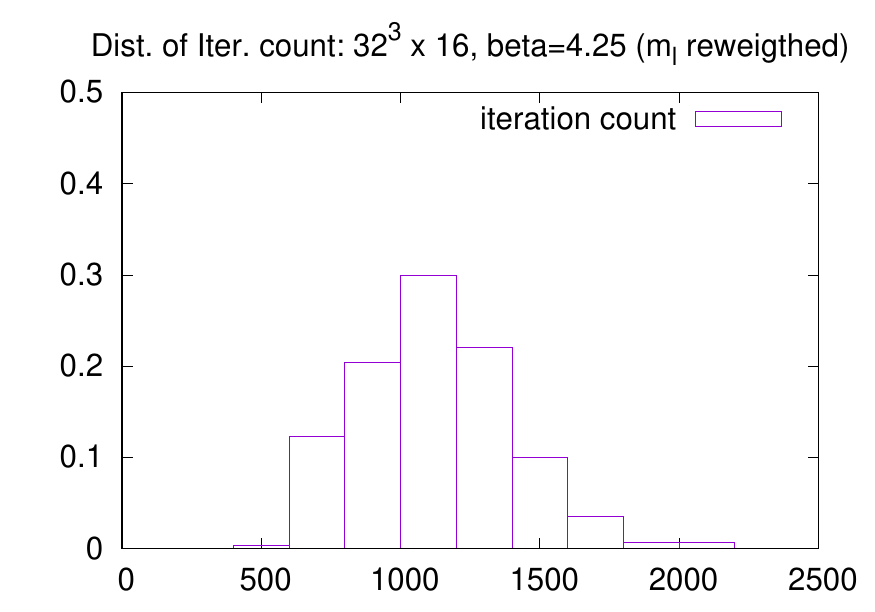}

\caption{
Distribution of the iteration counts of CG solver for the light quark during HMC, before (upper panels) and after (lower panels) reweighing for $N_T=16$ ensembles.  From the left panels to right: Temperature $T=129$ MeV ($\beta=4.10$), $153$ MeV ($\beta=4.17$) and $184$ MeV ($\beta=4.25$).}

\label{fig:reweight}

\end{figure}

We use the following three different lattices with $m_l = 0.1m_s$.
\begin{itemize}
 \item $24^3 \times 12$ lattice, 9 points with $\sim$ 10k -- 19k trajectories.
 \item $36^3 \times 12$ lattice, 8 points with $\sim$ 4k -- 12k trajectories.
 \item $32^3 \times 16$ lattice, 9 points with $\sim$ 4k -- 6k trajectories after mass reweighting ($\sim 20$k trajectories before reweighting).
\end{itemize}
Figure~\ref{fig:iter_l} shows iterations counts of CG solver for the light quark
during HMC and the variance,
which are not physical observables but sensitive to the low mode of the
Dirac spectrum.
Especially from the larger volume ($36^3\times 12$) data, we observe
that our choice of simulation points covers the (pseudo) transition point,
of which temperature is $150$ -- $170$ MeV.
The finer lattice data ($32^3 \times 16$) are mass reweighted from the
configurations generated without $m_{\mathrm{res}}$ corrections.
The distributions of the iteration counts before and after reweighting
are collected in Fig.~\ref{fig:reweight}.
The plots show that the distribution after the reweighting has a large
overlap with the original distribution, which justifies to use the reweighing method for this quantity\footnote{%
Post conference analysis with more statistics up to 20k trajectories gives smoother distribution than the figure.%
}.
We have also observed similar overlap for other quantities we present here.

\begin{figure}
 \centering
\includegraphics[width=0.48\linewidth]{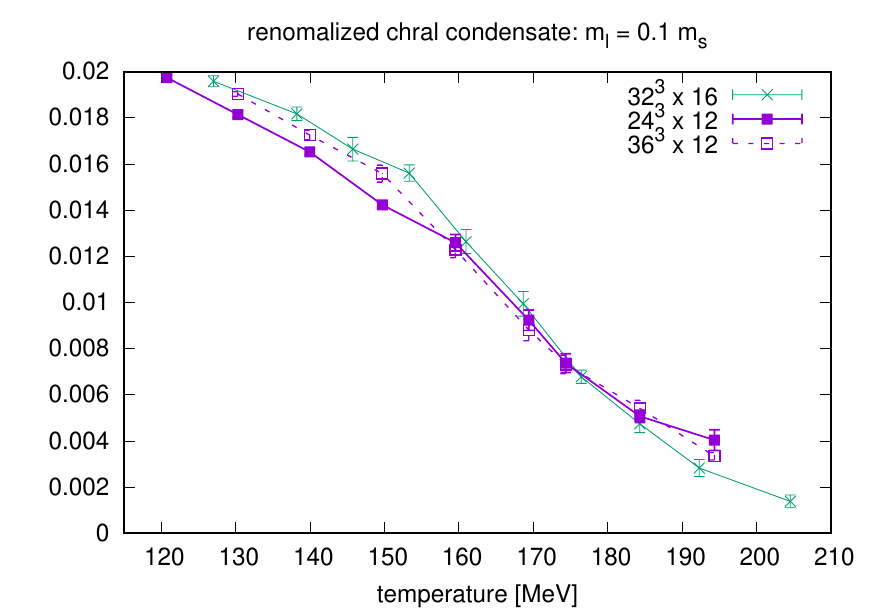}
\hfil
\includegraphics[width=0.48\linewidth]{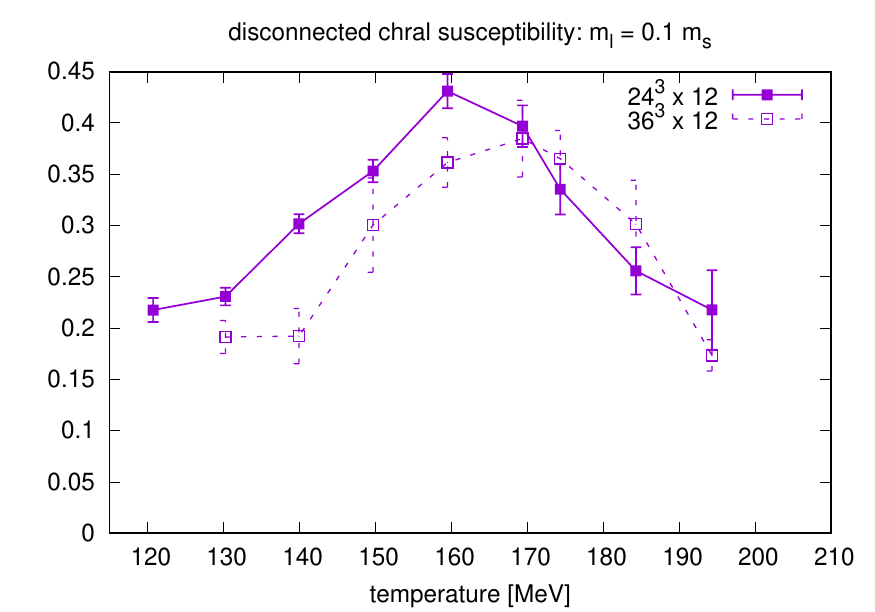}
\caption{Renormalized chiral condensate of the light quarks (left panel) and the disconnected part of the susceptibility (right panel).
The divergent part of the condensate is removed by subtracting the condensate of strange quark.  The multiplicative renormalization with $\mu=2$ GeV is also applied.  Mass reweighing is not applied to $N_T=16$ data.
}
\label{fig:condensate}
\end{figure}

Figure~\ref{fig:condensate} is the renormalized chiral condensate and disconnected
susceptibility.
Although the mass reweighting is not applied to the finer lattice data,
the values from two different lattice spacings, $N_T=12$ and $N_T=16$,
have almost the same value of chiral condensate.
This implies the divergent part of the condensate is properly subtracted
by using strange quark condensate as $\langle \bar\psi_l \psi_l\rangle - \frac{m_l}{m_s} \langle \bar\psi_s \psi_s\rangle $,
 and the multiplicative normalization ($\mu=2$ GeV) is properly applied.
The peak of the susceptibility is located in 150--170 MeV, but changes as the volume becomes larger.
We need a larger volume data to give a conclusive statement about the transition temperature.

\begin{figure}
\center
\includegraphics[width=0.47\linewidth]{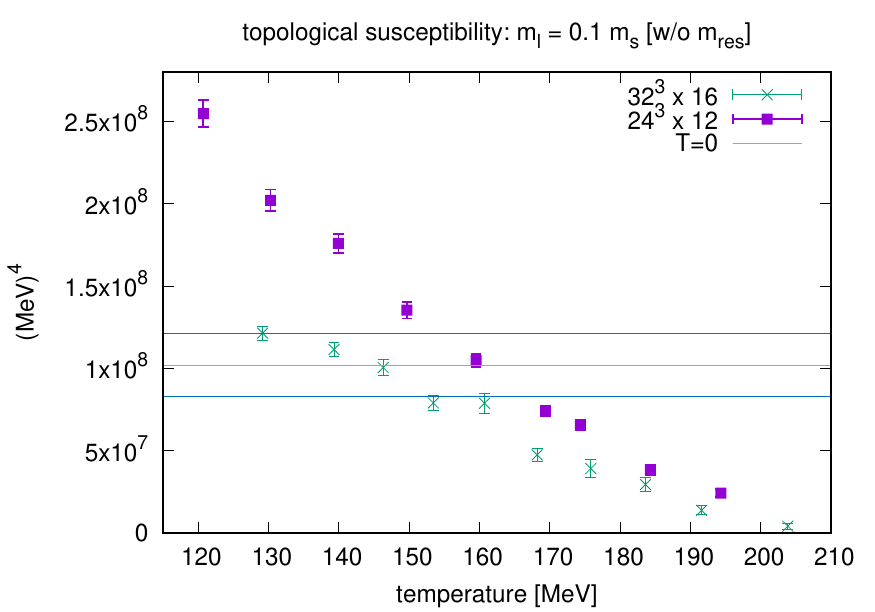}
\hfil
\includegraphics[width=0.47\linewidth]{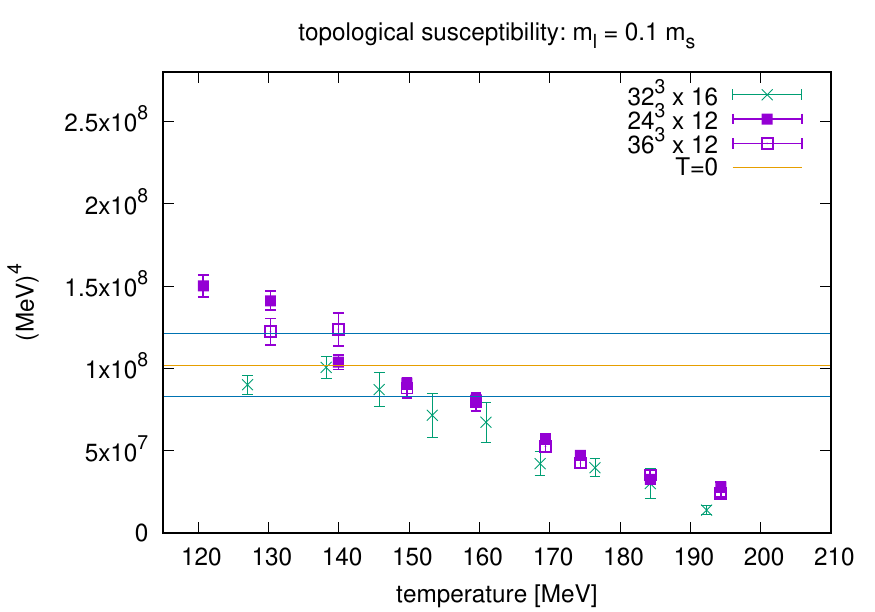}

\caption{Topological susceptibility measured by using gluonic definition.
The left panel is results without $m_{\mathrm{res}}$ correction and the right panel
is with $m_{\mathrm{res}}$ correction.
The horizontal lines denote the value at zero temperature and the error.
}
\label{fig:top}
\end{figure}

The topological susceptibility is rather sensitive to the quark mass.
Therefore, without the $m_{\mathrm{res}}$ correction we cannot
obtain the correct value.
The left panel of Fig.~\ref{fig:top} demonstrates this fact.
The plotted data with $N_T=12$ and $N_T=16$ differ only in the lattice spacing
but $m_{\mathrm{res}}$ corrections are not applied to both data series.
One would naively expect that they give almost the same value, however,
as the coarse lattice has much larger $m_{\mathrm{res}}$
the results significantly deviate from each other.  Furthermore, the value of the topological susceptibility with $N_T=12$ overshoots
the value at $T=0$ \cite{Aoki:2017paw} in the low temperature.
After the effect of $m_{\mathrm{res}}$ are corrected (right panel of Fig.~\ref{fig:top}), we observe a good agreement of $N_T=12$ and $N_T=16$ data in $T\gtrsim 140$~MeV.
In the lower temperature, we still observe an overshoots of $N_T=12$ data.
We interpret this is due to a remnant lattice artifact, 
because the lattice cut off of the lowest temperature simulation with $N_T=12$
is about 1.5 GeV, which is rather small.

\section{Summary and Outlooks}

We tuned the simulation parameters of 2+1 flavor 
configurations with M\"obius domain-wall fermions
to study the (pseudo) critical transition at finite temperature.
The simulation points are chosen along the line of constant physics,
where the light quark mass is set to $0.1 m_s$ in the physical unit.
The quark mass is tuned by taking into account the effect of the residual mass.
For the finer lattice with $N_T=16$, we used quark mass reweighting to correct
the residual mass effect (except for the chiral condensate of which reweighted results
are to come,  as additional measurements with the new valance quark mass are needed).
The preliminary results confirms the correctness of the parameters
and imply the transition temperature is in 150--170 MeV.

We are now adding more statistics and larger volume data, which 
are crucial to give a conclusive statements on the transition temperature
from our data.
The data with lighter quark mass at physical point are also to come.

\subsection*{Acknowledgments}

I.K. is supported by JSPS KAKENHI (JP20K03961) and, the MEXT as
`Program for Promoting Researches on the Supercomputer Fugaku' (Simulation for basic science:
from fundamental laws of particles to creation of nuclei) and `Priority Issue 9 to be Tackled
by Using the Post-K Computer' (Elucidation of The Fundamental Laws and Evolution of the Universe),
and Joint Institute for Computational Fundamental Science (JICFuS).
The simulations are performed on supercomputer ``Fugaku'' at RIKEN Center for Computational Science (HPCI project hp200130, hp210165, hp220174, and Usability Research ra000001), Oakforest-PACS at the Joint Center for Advanced High Performance Computing (JCAHPC) of the Universities of Tokyo and Tsukuba (hp200130), Polarie and Grand Chariot at Hokkaido University (hp200130).
We used code set Grid \cite{Grid, Boyle:2015tjk, Meyer:2021uoj} for configuration generations,
 a developing branch of Hadrons \cite{antonin_portelli_2020_4293902} and Bridge++ \cite{Bridge, Ueda:2014rya} for measurements.

\bibliographystyle{JHEP}
\bibliography{proc_lattice2022_finiteT_kanamori}

\providecommand{\href}[2]{#2}\begingroup\raggedright\begin{thebibliography}{10}

\bibitem{Aoki:2006we}
Y.~Aoki, G.~Endrodi, Z.~Fodor, S.D.~Katz and K.K.~Szabo, \emph{{The Order of
  the quantum chromodynamics transition predicted by the standard model of
  particle physics}}, \href{https://doi.org/10.1038/nature05120}{\emph{Nature}
  {\bfseries 443} (2006) 675}
  [\href{https://arxiv.org/abs/hep-lat/0611014}{{\ttfamily hep-lat/0611014}}].

\bibitem{Jin:2014hea}
X.-Y.~Jin, Y.~Kuramashi, Y.~Nakamura, S.~Takeda and A.~Ukawa, \emph{{Critical
  endpoint of the finite temperature phase transition for three flavor QCD}},
  \href{https://doi.org/10.1103/PhysRevD.91.014508}{\emph{Phys. Rev. D}
  {\bfseries 91} (2015) 014508}
  [\href{https://arxiv.org/abs/1411.7461}{{\ttfamily 1411.7461}}].

\bibitem{Jin:2017jjp}
X.-Y.~Jin, Y.~Kuramashi, Y.~Nakamura, S.~Takeda and A.~Ukawa, \emph{{Critical
  point phase transition for finite temperature 3-flavor QCD with
  non-perturbatively O($a$) improved Wilson fermions at $N_{\rm t}=10$}},
  \href{https://doi.org/10.1103/PhysRevD.96.034523}{\emph{Phys. Rev. D}
  {\bfseries 96} (2017) 034523}
  [\href{https://arxiv.org/abs/1706.01178}{{\ttfamily 1706.01178}}].

\bibitem{Bazavov:2017xul}
A.~Bazavov, H.T.~Ding, P.~Hegde, F.~Karsch, E.~Laermann, S.~Mukherjee et~al.,
  \emph{{Chiral phase structure of three flavor QCD at vanishing baryon number
  density}}, \href{https://doi.org/10.1103/PhysRevD.95.074505}{\emph{Phys. Rev.
  D} {\bfseries 95} (2017) 074505}
  [\href{https://arxiv.org/abs/1701.03548}{{\ttfamily 1701.03548}}].

\bibitem{Kuramashi:2020meg}
Y.~Kuramashi, Y.~Nakamura, H.~Ohno and S.~Takeda, \emph{{Nature of the phase
  transition for finite temperature $N_{\rm f}=3$ QCD with nonperturbatively
  O($a$) improved Wilson fermions at $N_{\rm t}=12$}},
  \href{https://doi.org/10.1103/PhysRevD.101.054509}{\emph{Phys. Rev. D}
  {\bfseries 101} (2020) 054509}
  [\href{https://arxiv.org/abs/2001.04398}{{\ttfamily 2001.04398}}].

\bibitem{Dini:2021hug}
L.~Dini, P.~Hegde, F.~Karsch, A.~Lahiri, C.~Schmidt and S.~Sharma,
  \emph{{Chiral phase transition in three-flavor QCD from lattice QCD}},
  \href{https://doi.org/10.1103/PhysRevD.105.034510}{\emph{Phys. Rev. D}
  {\bfseries 105} (2022) 034510}
  [\href{https://arxiv.org/abs/2111.12599}{{\ttfamily 2111.12599}}].

\bibitem{Cuteri:2021ikv}
F.~Cuteri, O.~Philipsen and A.~Sciarra, \emph{{On the order of the QCD chiral
  phase transition for different numbers of quark flavours}},
  \href{https://doi.org/10.1007/JHEP11(2021)141}{\emph{JHEP} {\bfseries 11}
  (2021) 141} [\href{https://arxiv.org/abs/2107.12739}{{\ttfamily
  2107.12739}}].

\bibitem{Colquhoun:2022atw}
{\scshape JLQCD} collaboration, \emph{{Form factors of
  B\textrightarrow{}\ensuremath{\pi}\ensuremath{\ell}\ensuremath{\nu} and a
  determination of |Vub| with M\"obius domain-wall fermions}},
  \href{https://doi.org/10.1103/PhysRevD.106.054502}{\emph{Phys. Rev. D}
  {\bfseries 106} (2022) 054502}
  [\href{https://arxiv.org/abs/2203.04938}{{\ttfamily 2203.04938}}].

\bibitem{talk_yaoki}
Y.~Aoki, S.~Aoki, F.~Hidenori, K.I.~Hashimoto, Shoji, T.~Kaneko and
  Y.~Nakamura, Yoshifumi an~Zhang, ``{Thermodynamics with Möbius domain wall
  fermions near physical point (I)}.'' A talk at this conference.

\bibitem{Aoki:2017paw}
{\scshape JLQCD} collaboration, \emph{{Topological susceptibility of QCD with
  dynamical M\"obius domain-wall fermions}},
  \href{https://doi.org/10.1093/ptep/pty041}{\emph{PTEP} {\bfseries 2018}
  (2018) 043B07} [\href{https://arxiv.org/abs/1705.10906}{{\ttfamily
  1705.10906}}].

\bibitem{Grid}
``{Grid}.'' \url{https://github.com/paboyle/Grid}.

\bibitem{Boyle:2015tjk}
P.~Boyle, A.~Yamaguchi, G.~Cossu and A.~Portelli, \emph{{Grid: A next
  generation data parallel C++ QCD library}},
  \href{https://arxiv.org/abs/1512.03487}{{\ttfamily 1512.03487}}.

\bibitem{Meyer:2021uoj}
N.~Meyer, P.~Georg, S.~Solbrig and T.~Wettig, \emph{{Grid on QPACE 4}},
  \href{https://doi.org/10.22323/1.396.0068}{\emph{PoS} {\bfseries LATTICE2021}
  (2022) 068} [\href{https://arxiv.org/abs/2112.01852}{{\ttfamily
  2112.01852}}].

\bibitem{antonin_portelli_2020_4293902}
A.~Portelli, N.~Asmussen, P.~Boyle, F.~Erben, V.~Gülpers, R.~Hodgson et~al.,
  \emph{aportelli/hadrons: Hadrons v1.2},  Nov., 2020.
\newblock 10.5281/zenodo.4293902.

\bibitem{Bridge}
``{Lattice QCD code Bridge++}.''
  \url{https://bridge.kek.jp/Lattice-code/index_e.html}.

\bibitem{Ueda:2014rya}
S.~Ueda, S.~Aoki, T.~Aoyama, K.~Kanaya, H.~Matsufuru, S.~Motoki et~al.,
  \emph{{Development of an object oriented lattice QCD code 'Bridge++'}},
  \href{https://doi.org/10.1088/1742-6596/523/1/012046}{\emph{J. Phys. Conf.
  Ser.} {\bfseries 523} (2014) 012046}.

\end{thebibliography}\endgroup

\end{document}